\documentclass[aps,prb,twocolumn,showpacs]{revtex4}
\usepackage{graphics}
\usepackage[caption=false]{subfig}
\usepackage{epsfig}
\usepackage{color}
\usepackage{amsmath}
\usepackage{amssymb}
\usepackage{amsfonts}
\usepackage{wrapfig}
\usepackage{pstricks}
\usepackage{pst-node}
\usepackage{bm}
\usepackage{dcolumn}
\usepackage{multirow}

\begin{document}
\title{Strain effects on the spin-orbit induced band structure splittings in monolayer MoS$_2$ and graphene.}
\author{Tawinan Cheiwchanchamnangij and Walter R. L. Lambrecht}
\affiliation{Department of Physics, Case Western Reserve University,
Cleveland, OH 44106-7079}
\author{Yang Song$^1$ and Hanan Dery$^{1,2}$}
\affiliation{$^1$Department of Physics and Astronomy,
$^2$ Department of Electrical and Computer Engineering,
University of Rochester, Rochester, New York, 14627}
\begin{abstract}
The strain effects on the spin-orbit induced splitting of the valence band
maximum and conduction band minimum in monolayer MoS$_2$ and the gap in graphene
are calculated using first-principles calculations. The dependence of these
splittings on the various symmetry types of strain is described by means of an
effective Hamiltonian based on the method of invariants and the parameters
in the model are extracted by fitting to the theory. These splittings
are related to acoustic phonon deformation potentials, or electron-phonon
coupling matrix elements which enter the spin-dependent
scattering theory of conduction in these materials.
\end{abstract}
\maketitle

\section{Introduction}
Graphene and monolayer MoS$_2$ are both interesting materials for
spin-dependent electronic devices.\cite{Dery_IEEE12,Gong_NatureComm13} In spite of their similarities,
they also have significant differences. As is well known,
graphene has a linear dispersion near the Dirac points and has inversion
symmetry. MoS$_2$ has a gap of about 1.8 eV and the absence of
inversion symmetry in monolayer MoS$_2$ leads to an interesting relation
of the spins and the valley degrees of freedom.
Both valence band and conduction band
edges in the $K$ and $K'$ valleys of the Brillouin zone
are split purely by spin-orbit coupling. Because of the time-reversal
symmetry between $\psi_{{\bf k}\sigma}$ and $\psi_{-{\bf k}-\sigma}$, the
up and down spin states in opposing valleys are reversed. This relation leads
to the possibility of valley control of the carriers by means of
circularly polarized excitation.\cite{Feng2012,Mak2012,Zeng2012,Kioseoglou_APL12,Sallen_PRB12}
On the other hand, spin transport in this material depends on the intra-valley
and inter-valley scattering, which arises from the
electron-phonon coupling. As is also well known in graphene, ripples play an important role
in 2D materials.\cite{Castro_RMP09} These ripples are governed by out-of-plane long-wavelength phonon distortion; the so-called flexural acoustic mode.
This mode is associated with the dynamic out-of-plane shear strain. Thus, by studying
the splittings and shifts of the energy bands with strain, it is possible to
extract deformation potential constants that set the amplitudes of various electron-phonon scattering processes.\cite{Song_PRL13,Kaasbjerg_PRB12,Kaasbjerg_PRB13}
The present study  is motivated by this connection.

Strain induced changes in the band gap of MoS$_2$ have been the subject
of several recent papers.\cite{Yun12,Scalise12,Lu12,Pan12,Yue12,Peelaers12}
 The emphasis of those papers is on the
strain engineering of the band structure. While we will also present some
results on the band gaps with strain, our emphasis is on studying the effect of different
types of strain on the band edge spin-orbit splittings in MoS$_2$.
Closely related, the spin-orbit coupling in graphene leads to the opening
of a gap at the Dirac point.\cite{Kane05}
The size of this spin-orbit coupling induced
gap in graphene is extremely small and has been somewhat controversial with different
estimates resulting from different tight-binding models and
first-principles calculations.\cite{Min06,Huertas06,Gmitra09,Konschuh10}  Here, we  present first-principles calculations
of this spin-orbit induced gap, and from its dependence on strain we recover the scattering constant for intrinsic spin flips in graphene.\cite{Song_PRL13}
This connection is made via a new strain term that we introduce to the $K$-point effective Hamiltonian in graphene.

In the theory of electron-phonon scattering, absolute deformation potentials
of individual bands play a role.\cite{Bir_Pikus_Book,Cardona,Song_PRB12}
 The so-called absolute hydrostatic deformation potentials lead to shifts of the bands and cannot be extracted from
a single bulk calculation because the reference potential in a
periodic crystal is ill-defined.\cite{Kleinman81,Cardona87,VandeWalle89}
 An interface calculation is needed
between a strained and unstrained region in order to obtain the
dipole potential alignment between the two and hence the absolute shifts
of the bands. On the other hand, splitting of the bands due to
traceless components of the strain tensor are obtainable from a single
unit cell calculation, appropriately strained. In  this paper, we focus
only on deformation potentials which can be extracted from band splittings.
It also means that we cannot obtain inter-valley matrix elements in this manner.
In spite of these restrictions, the present study should be of interest,
because the splittings studied under strain are in principle also directly
observable experimentally.

Before embarking on the theory we note that it may be surprising that the spin-orbit
splitting would depend on strain at all. After all, the spin-orbit coupling
is a relativistic effect resulting mostly from the inner parts of the atom
and hence be mostly an atomic property. However, the strain affects the mixing
of different atomic orbitals (e.g., different Mo-$d$ orbitals and S-$p$ orbitals in the eigentstates), and as a result the effective splitting does depend
measurably on strain. While these effects are indeed rather small, they
present a challenge to the computational accuracy.  Nevertheless, systematic errors of density functional theory in its
usual local density approximation cancel out in these energy differences (band splittings)
and differences of differences  (strain induced changes in the splittings). This fact
makes it possible to calculate the splittings as long as a sufficiently  accurate basis set is used to find the eigenvalues of the crystal potential (i.e.,
the band structure). The all-electron linearized band structure methods like the full-potential
linearized muffin-tin orbital method (FP-LMTO) satisfies the requirements.

The plan of the paper is as follows. In Sec.~\ref{theory}
we derive effective Hamiltonian forms using the method of invariants.
In Sec.~\ref{method} we present details of the first-principles
computational method employed. In Sec.~\ref{results}, we present
the first-principles results on the splittings under the effect of different
types of strain. These results confirm the strain dependence of the splittings on strain as predicted by the effective Hamiltonian models. The parameters of these models are then obtained by fitting the strain dependence to the calculated curves.
For the importance of these parameters in the description of
the spin-dependent relaxation processes, we refer the reader to Ref.~[\onlinecite{Song_PRL13}].

\section{Theory}\label{theory}

The dependence of spin-orbit coupling induced splitting on strain
is governed by an effective Hamiltonian describing only the states
near this splitting as function of different strain components. To obtain the forms
of these effective Hamiltonians, we employ the method of invariants.\cite{Luttinger_PR56,Bir_Pikus_Book,Winkler_Book,Winkler_PRB10,Song_PRB12} This  group theoretical framework
provides the terms allowed by symmetry. Simply put, the strain tensor is
decomposed in irreducible representations of the point group of the
${\bf k}$-point where the band splitting of interest occurs.
The Hamiltonian must belong to the fully symmetric irreducible representation (IR).
Group theory thus determines which terms, linear or quadratic in the
strain component, are allowed in the Hamiltonian.
In the case of MoS$_2$, the states near the splitting are determined
by a $2\times2$ Hamiltonian while in the case of graphene, the Dirac cone
is represented by a $4\times4$ Hamiltonian, including both the orbital
pseudospin and real spin degrees of freedom. The theory thus predicts
linear or quadratic dependence of the band splittings on different
types of strain.

\subsection{MoS$_2$ effective Hamiltonian}\label{theory1}
In the absence of strain, the spin-orbit split states at the $K$-point band edges in MoS$_2$ are described by a $2\times2$ matrix, conveniently
expressed in terms of the Pauli matrices and unit matrix:
\begin{equation}
H_0=\bar E {\bf 1}+ \frac{1}{2}\Delta_0\sigma_z.
\end{equation}
Here $\bar E$ is the average of up and down spin bands without spin-orbit splitting and as an arbitrary constant it can be set to zero. $\Delta_0$ is the
spin-orbit splitting.  The same form applies to the valence band maximum
(VBM) and conduction band minimum (CBM) although of course the value of
$\Delta_0$ is rather different.\cite{Cheiwchanchamnangij} We distinguish them by means of a subscript
$v$ or $c$ for VBM and CBM, respectively.
It is much smaller for the CBM ($\Delta_{0c}=3.36$~meV)
than for the VBM ($\Delta_{0v}=146$~meV) because the CBM
states are derived mostly from the Mo-$d_{3z^2-r^2}$ orbital which has
quantum number $L_z=0$, and hence zero spin-orbit coupling.
It is only because of the small deviation of the Mo
atomic potential from spherical
symmetry due to the crystal structure, and because of the small components
from S $p_x,p_y$ orbitals that it is not zero.\cite{Zhu_PRB11}
On the other hand, the VBM consists of $d_{x^2-y^2},d_{xy}$ like states.
The strain adds terms $b(\bm{\epsilon})$
on the diagonal and $a_i(\bm{\epsilon})$ $(i=x,y)$
on the off-diagonal of the form
\begin{equation} \label{eq:Hstr}
H_{strain}=\frac{1}{2}b(\bm{\epsilon})\sigma_z+ \frac{1}{2}\sum_{i=x,y}a_i(\bm{\epsilon})\sigma_i.
\end{equation}
We employ symmetrized strain-tensor components, $\epsilon_{ij}=(du_i/dx_j+du_j/dx_i)/2$, where $u_i$ is the displacement
of the $i$-th cartesian coordinate.\footnote{In Ref.~[\onlinecite{Song_PRL13}] we worked with non-symmetrized forms in which out-of-plane displacements ($du_z/dx_i$) are treated separately. Therefore, the scattering parameter $\Xi_{K}^{so}$ in Ref.~[\onlinecite{Song_PRL13}] is to be associated with $a_i$/4 in this work [Eq.~(\ref{eq:Hstr})].} The strain dependent splitting of $H_0 + H_{strain}$ becomes
\begin{equation}
\Delta (\bm{\epsilon})=\sqrt{[\Delta_0+b(\bm{\epsilon})]^2+|a(\bm{\epsilon})|^2}.
\end{equation}
Next, we specify the dependence of $b(\bm{\epsilon})$ and $a(\bm{\epsilon})$
for different symmetry components of the strain tensor.
The flexural modes or out-of-plane
strain-tensor components ($\epsilon_{xz}$ and $\epsilon_{yz}$) are odd with respect to the mirror-plane passing through the
MoS$_2$ layer. Accordingly, they can only contribute even terms to the diagonal and odd terms
to the off-diagonal parts of the effective Hamiltonian. Hence, to lowest
non-zero order,
\begin{eqnarray} \label{eq:b_and_a}
a_i(\epsilon_{iz}) \approx a_1\epsilon_{iz}\,, \qquad \qquad b(\epsilon_{iz}) \approx b_2\epsilon_{iz}^2.
\end{eqnarray}
Contrary to the out-of-plane case, in-plane strain components ($\epsilon_{xx}, \epsilon_{yy}, \epsilon_{xy}$)
are even with respect to the horizontal mirror-plane. The off-diagonal
terms must necessarily be zero and cannot lead to spin-flip scattering,\cite{Song_PRL13}
whereas the leading diagonal term is linear in strain,
\begin{eqnarray}
a_i(\epsilon_{ij})=0\,, \qquad \,\,\,\,\, b(\epsilon_{ij}) \approx b_1\epsilon_{ij}+b_2\epsilon_{ij}^2+\dots.
\end{eqnarray}
For the VBM under out-of-plane strain $\epsilon_{iz}$, the zero strain
splitting is large compared to the strain terms and one might thus
think that the term $b_{2v}\epsilon_{iz}^2$ may be entirely neglected. In that case
expanding the square root, one obtains
\begin{equation}
 \Delta_v(\epsilon_{iz})=\Delta_{0v}+\frac{a_{1v}^2\epsilon_{i,z}^2}{2\Delta_{0v}},
\end{equation}
which predicts a quadratic increase of the spin-orbit splitting
with strain. However, we find that the splitting
actually decreases. This implies that $b_{2v}\epsilon_{iz}^2$ cannot be neglected
and has negative $b_{2v}$ because it is the only contribution that can reduce the splitting.
So keeping this important term, we can write
\begin{equation}
\Delta_v(\epsilon_{iz})^2=\Delta_{0v}^2+(a_{1v}^{2}+2b_{2v}\Delta_0)\epsilon_{iz}^2+b_{2v}^2\epsilon_{iz}^4.
\label{eqspliteiz}
\end{equation}
For the CBM and out-of-plane strain, the zero-strain splitting is
much smaller so that the strain may become dominant and give rise to a
linear strain dependence,
\begin{equation}
\Delta_c(\epsilon)\approx |a_{1c}|\epsilon_{iz}
\end{equation}
except for very small strains, where it should still
look quadratic.

Turning to in-plane strains, there are no off-diagonal terms and the leading term is
linear in strain, so we simply have
\begin{equation}
\Delta_c(\epsilon_{ij})=\Delta_{0c}+b_{1c}\epsilon_{ij}+b_{2c}\epsilon_{ij}^2. \label{inplane}
\end{equation}
However, the $b_{1c}$ for $\epsilon_{xy}$ vanishes, in which case the dependence is
again quadratic on strain. In Appendix \ref{app:b1}, we derive this exact cancelation of $b_{1c}$ for $\epsilon_{xy}$, and also the analytical expression of $b_{1c}$ for $\epsilon_{ii}$. The latter is shown to result from competition between first and second order perturbation terms. For the conduction band, where the spin-orbit splitting is small,
the linear in strain term turns out to be  also small for tensional strain in the plane.
Thus, the quadratic term can become dominant for sufficiently large
strains. We will see that in fact $b_{2c}$ is negative in that case.

\subsection{Graphene effective Hamiltonian}\label{theory2}
To write the strain-dependent Hamiltonian in graphene, we invoke the transformation properties of the states and strain tensor in its $K$~point. This information is summarized in  Table~\ref{tab:CharactorTable_D_3h} (Appendix~\ref{app:b2}). The states in the edges of the conduction and valence bands of the Dirac point transform as $\Gamma_7$ and $\Gamma_9$. Each of these two-dimensional IRs reflects the sublattice orbital degeneracy (pseudospin) and real spin degeneracy due to the space inversion symmetry.\cite{Winkler_PRB10} Table~\ref{tab:CharactorTable_D_3h} also includes invariants that tell us that the coupling between the edge states and strain can be written in the following Hamiltonian form
\begin{equation}
\!\!H\!=\!\left(\!\!\begin{array}{cccc}
a(\bm{\epsilon})\!+\!\frac{\Delta_D}{2}&0&b(\bm{\epsilon})& ic(\bm{\epsilon}) \\
0& a(\bm{\epsilon})\!+\!\frac{\Delta_D}{2}&ic^*(\bm{\epsilon})&b^*(\bm{\epsilon})\\
b^*(\bm{\epsilon})&-ic(\bm{\epsilon})& a(\bm{\epsilon})\!-\!\frac{\Delta_D}{2}& 0 \\
-ic^*(\bm{\epsilon})&b(\bm{\epsilon})& 0 & a(\bm{\epsilon})\!-\!\frac{\Delta_D}{2}
\end{array}\!\!\right)\!.\! \label{eq:H_strain_graphene}
\end{equation}
$\Delta_D$ is the strain independent gap induced by spin-orbit coupling.\cite{Kane05} This gap separates the edge states of the conduction and valence bands in the Dirac point.  The diagonal strain term, $a(\bm{\epsilon})$, is an hydrostatic deformation potential that merely shifts the Dirac point relative to the average electrostatic potential. Its magnitude scales with the local changes in the density of electrons in response to contraction or dilatation of the unit-cell.\cite{Castro_RMP09,Suzuura_PRB02} The connection with strain components follows the transformation properties of the identity IR ($\Gamma_1$),
\begin{eqnarray}
a(\bm{\epsilon})&=&\Xi_d \left[ (\epsilon_{xx}+\epsilon_{yy}) - \tfrac{1}{2}(\epsilon_{xz}^2+\epsilon_{yz}^2) \right]. \label{eq:str_diag_a}
\end{eqnarray}
The hydrostatic term comprises linear (quadratic) components of in-plane (out-of-plane) strain. These terms share a single deformation potential constant ($\Xi_d$) because graphene has only one atomic layer; unlike MoS$_2$ which has some intrinsic thickness, in undistorted graphene all atoms lie in the same plane. This fact means that a flexural strain ($\epsilon_{xz}, \epsilon_{yz}$) can be viewed as a mere stretch of the membrane. Using elementary geometry, it is readily seen that $\epsilon_{xz}^2/2$ or $\epsilon_{yz}^2/2$ correspond to $\epsilon_{xx}$ or $\epsilon_{yy}$. To represent this physics, we also use $\epsilon_{iz} = \partial u_z/\partial x_i$ to define the out-of-plane tensor component in graphene.

Contrary to the diagonal strain term, the off-diagonal terms in Eq.~(\ref{eq:H_strain_graphene}) affect the size of the gap according to
\begin{equation}
E_g=\sqrt{\Delta_D^2+4|b(\bm{\epsilon})|^2+4|c(\bm{\epsilon})|^2}.
\end{equation}
$b(\bm{\epsilon})$ and $c(\bm{\epsilon})$ are shear-strain components that couple the edge states ($\Gamma_7 \times \Gamma_9 = \Gamma_5 + \Gamma_6$). The coupling between states of similar spin possesses $\Gamma_6$ symmetry, and its form reads
\begin{eqnarray}
b(\bm{\epsilon})=\Xi_o\left[\epsilon_{xx}-\epsilon_{yy}+2i\epsilon_{xy}-\frac{\epsilon_{xz}^2-\epsilon_{yz}^2+2i\epsilon_{xz}\epsilon_{yz}}{2}\right]\!.\,\,\,\, \label{eq:str_non_diag_b}
\end{eqnarray}
Again, we may see that this coupling comprises linear (quadratic) components of in-plane (out-of-plane) strain. The spin-independent deformation potential constant, $\Xi_o$, is often described by means of a fictitious vector potential due to changes in the hopping energy between nearest neighbor orbitals.\cite{Castro_RMP09,Suzuura_PRB02,Kane_PRL97,Castro10} Importantly, as shear strain does not change the unit-cell area to leading order, this parameter is not associated with local changes in the density of electrons. Hence, it can be calculated via density functional theory in the local density approximation.

Compared with previous works,\cite{Castro_RMP09,Suzuura_PRB02,Kane_PRL97,Castro10} our contribution to the strain Hamiltonian of graphene is the spin-orbit coupling term in the anti-diagonal of Eq.~(\ref{eq:H_strain_graphene}). It possesses $\Gamma_5$ symmetry due to the coupling between states of opposite spin, and its form reads
\begin{eqnarray}
c(\bm{\epsilon})&=& c_{so}(\epsilon_{yz}+i\epsilon_{xz}).
\end{eqnarray}
Contrary to the spin-independent terms, this term comprises linear out-of-plane strain. We will make use of this feature to extract its magnitude. The importance of the spin-orbit coupling deformation potential constant, $c_{so}$, is realized from the fact that it is directly related to the intrinsic spin relaxation in graphene.\footnote{The relation between $c_{so}$ in this work and the scattering constant $\Xi_{so}$ in Eq.~(6) of Ref.~[\onlinecite{Song_PRL13}] is $\Xi_{so}=2c_{so}$.}

Putting these pieces together, the change of the gap in response to pure in-plane strain becomes
\begin{equation}
E_g=\sqrt{\Delta_D^2+4\Xi_o^2[(\epsilon_{xx}-\epsilon_{yy})^2+4\epsilon_{xy}^2]}. \label{eq:Eg_gr_inplane}
\end{equation}
Similarly, the change of the gap in response to pure out-of-plane strain becomes
\begin{equation}
E_g\!=\!\sqrt{\Delta_D^2 + \Xi_o^2\left(\epsilon_{xz}^2+\epsilon_{yz}^2\right)^2 + 4c_{so}^2(\epsilon_{yz}^2+\epsilon_{xz}^2)}.  \label{eq:Eg_gr_outplane}
\end{equation}
In the following, we will use Eqs.~(\ref{eq:Eg_gr_inplane})~and~(\ref{eq:Eg_gr_outplane}) to determine the corresponding shear deformation potential parameters $\Xi_o$ and $c_{so}$ by fitting to first-principles calculations including appropriate strain combinations.

\section{Computational Method}\label{method}
Density functional theory is used in the local density approximation
following the von-Barth-Hedin parametrization. The band structures are calculated
using the full-potential linearized muffin-tin orbital (FP-LMTO) method
as described in Refs.~[\onlinecite{Methfessel}]~and~[\onlinecite{Kotani10}].  A double ($\kappa$,$R_{sm}$)
basis set is used, including $spdf$ and $spd$ for the first  and
second set respectively. Here, $\kappa^2=E-v_{mtz}$ represents the
kinetic energy of the smoothed Hankel function, or its decay length, while the
$R_{sm}$ is a smoothing radius (See Bott \textit{et al.} in Ref.~[\onlinecite{Bott}]).
Brillouin zone integrations are carried out using a 13$\times$13$\times$3 and 14$\times$14$\times$7 k-point sets for graphene and MoS$_2$, respectively.
Augmentation is carried out inside the muffin-tin-spheres up to $l_{max}=4$.

The LDA underestimates the gap significantly
in MoS$_2$,\cite{Cheiwchanchamnangij} but for the
changes in the splitting considered here, it is sufficient to consider the
quasiparticle self-energy shift to be indendent of strain.
Similarly, in graphene the LDA underestimates the Fermi velocity or
slope of the Dirac cone. Using the quasiparticle-self-consistent $GW$ method,\cite{kotani:QSGW} we obtain a change of the Dirac cone slope from
0.8$\pm$0.1$\times10^6$ m/s in LDA to 1.1$\pm$0.1$\times10^6$ m/s in good agreement with
van Schilfgaarde and Katsnelson.\cite{markgraphene}
For the present purposes of deriving the strain dependent deformation potentials,
the LDA results are deemed sufficiently accurate.

\section{Results}\label{results}
\subsection{MoS$_2$}
\begin{figure}
	\subfloat[]{\label{exz_vk}\includegraphics[width=75mm]{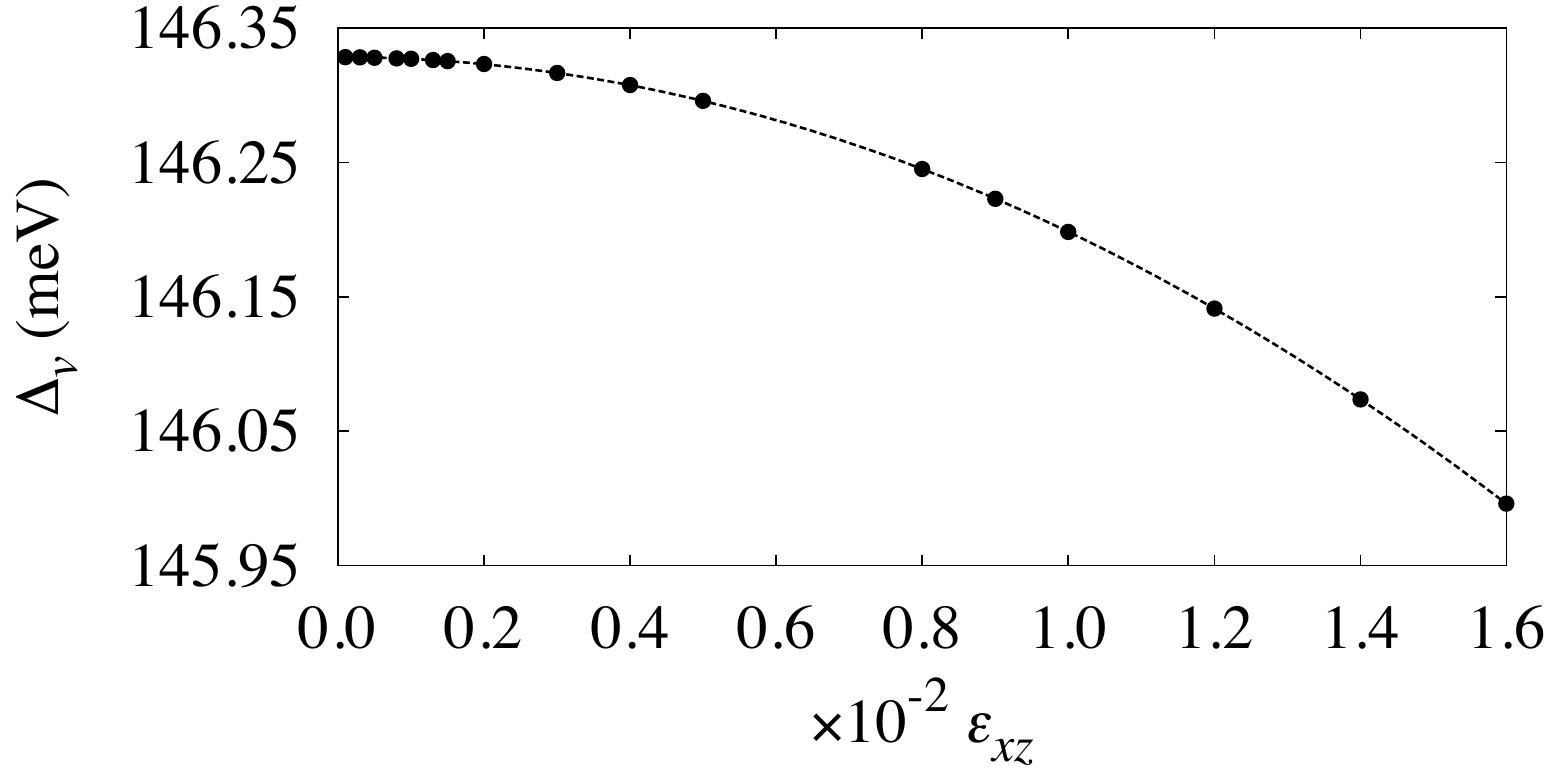}}
	
	\subfloat[]{\label{exz_ck}\includegraphics[width=75mm]{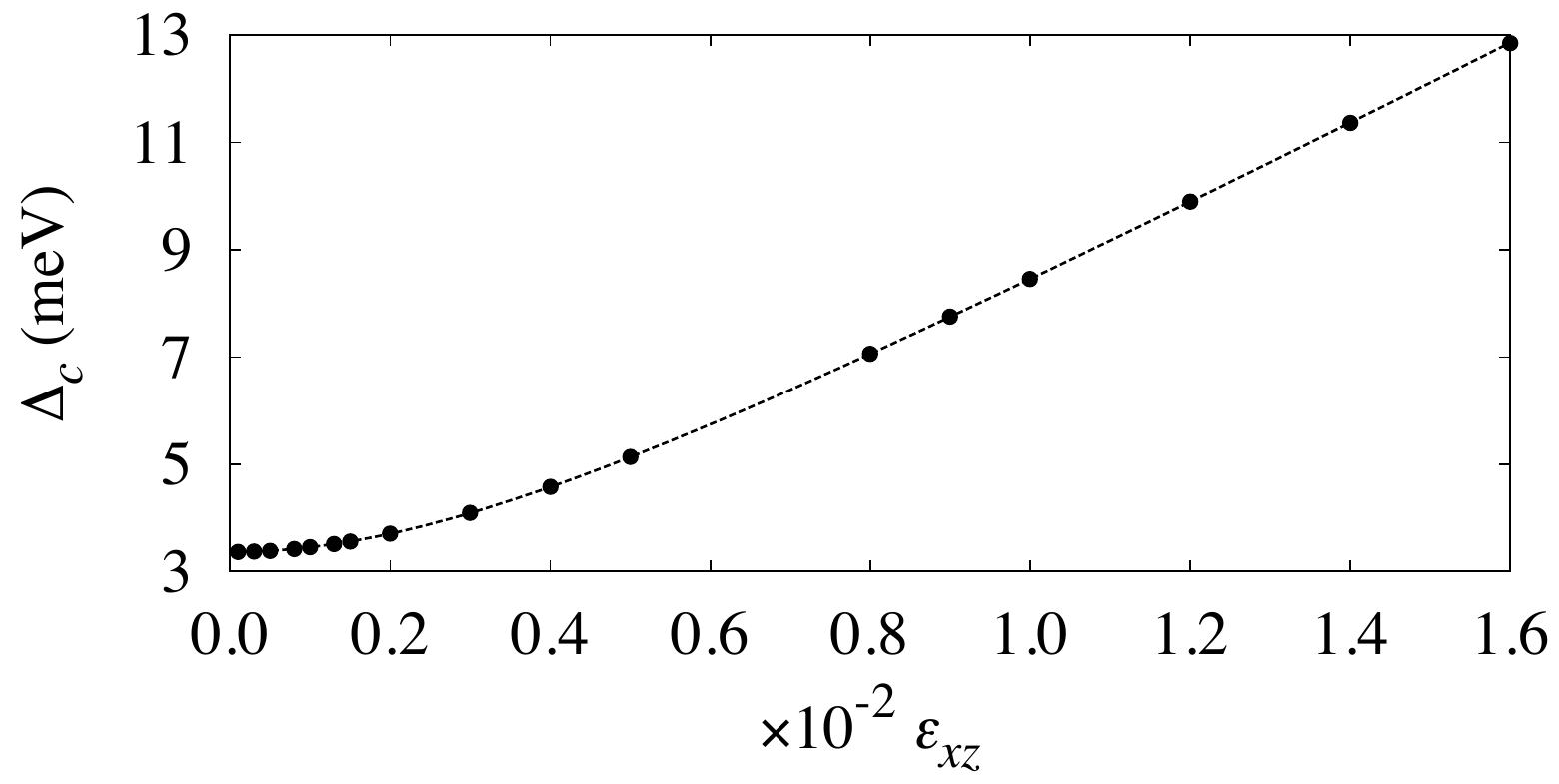}}
	
	
	
	\subfloat[]{\label{exz_gap}\includegraphics[width=75mm]{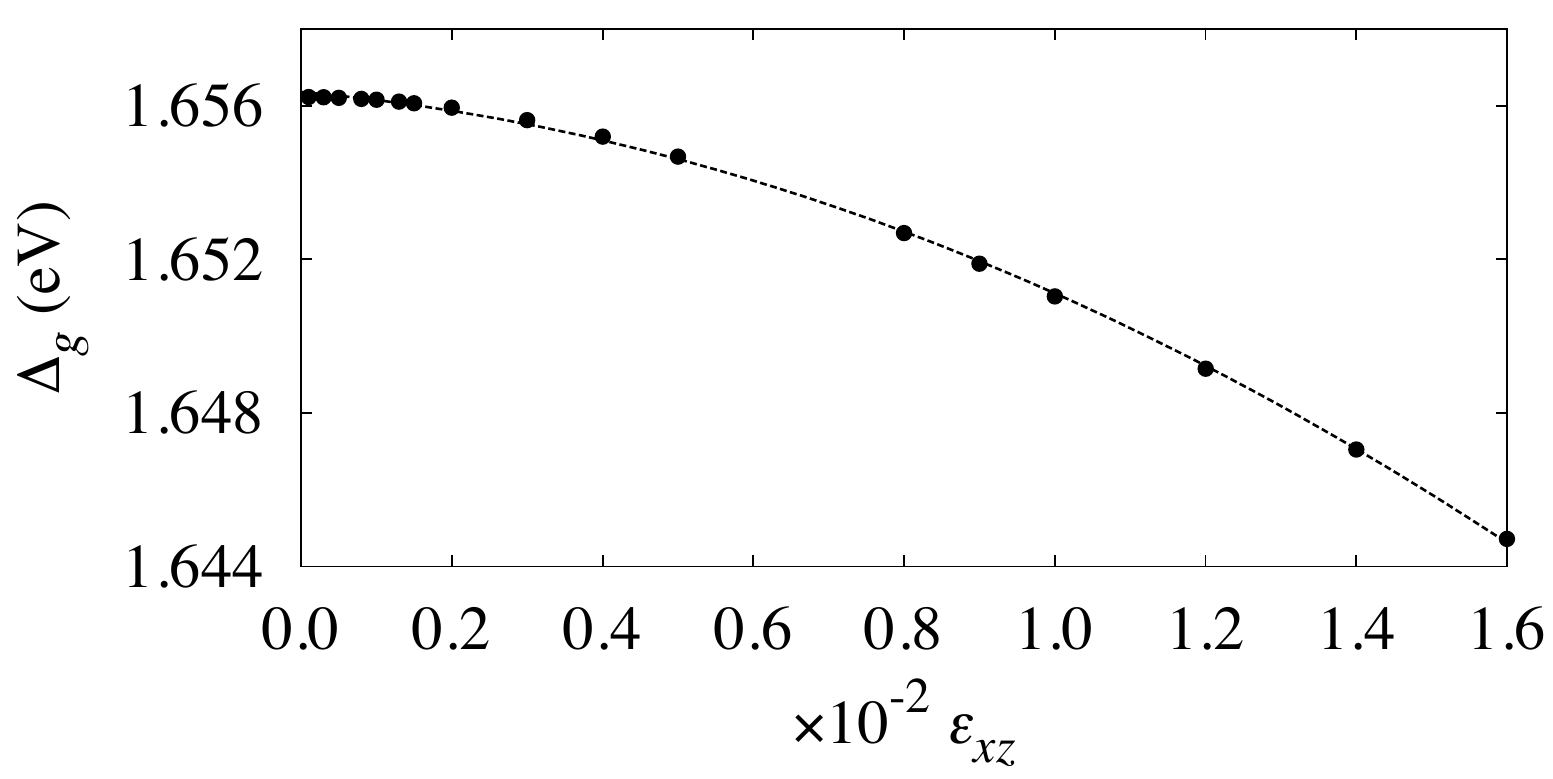}}
	
	\caption{(a) valence band splitting, (b) conduction band splitting, and (c) band gap of monolayer MoS$_2$ as a function of out-of-plane strain ($\epsilon_{xz}$)}
	\label{exz}
\end{figure}

Figure~\ref{exz_vk} shows  the splitting of the valence band
as a function of out-of-plane strain. As already mentioned, the splitting decreases
as function of strain. The fitted parameters to Eq.~(\ref{eqspliteiz}) are $a_{1v}=573$ meV and  $b_{2v}=-2.42$ eV.
Figure~\ref{exz_ck} shows the splitting of the conduction band as a function of
out-of-plane strain. We can see that it behaves nearly linear for sufficiently large strain. The parameters were obtained by directly fitting the square-root behavior and resulted in $a_{1c}=774$ meV and $b_{2c}=263$ meV.
\begin{figure}[h]
	\subfloat[]{\label{exx_vk}\includegraphics[width=75mm]{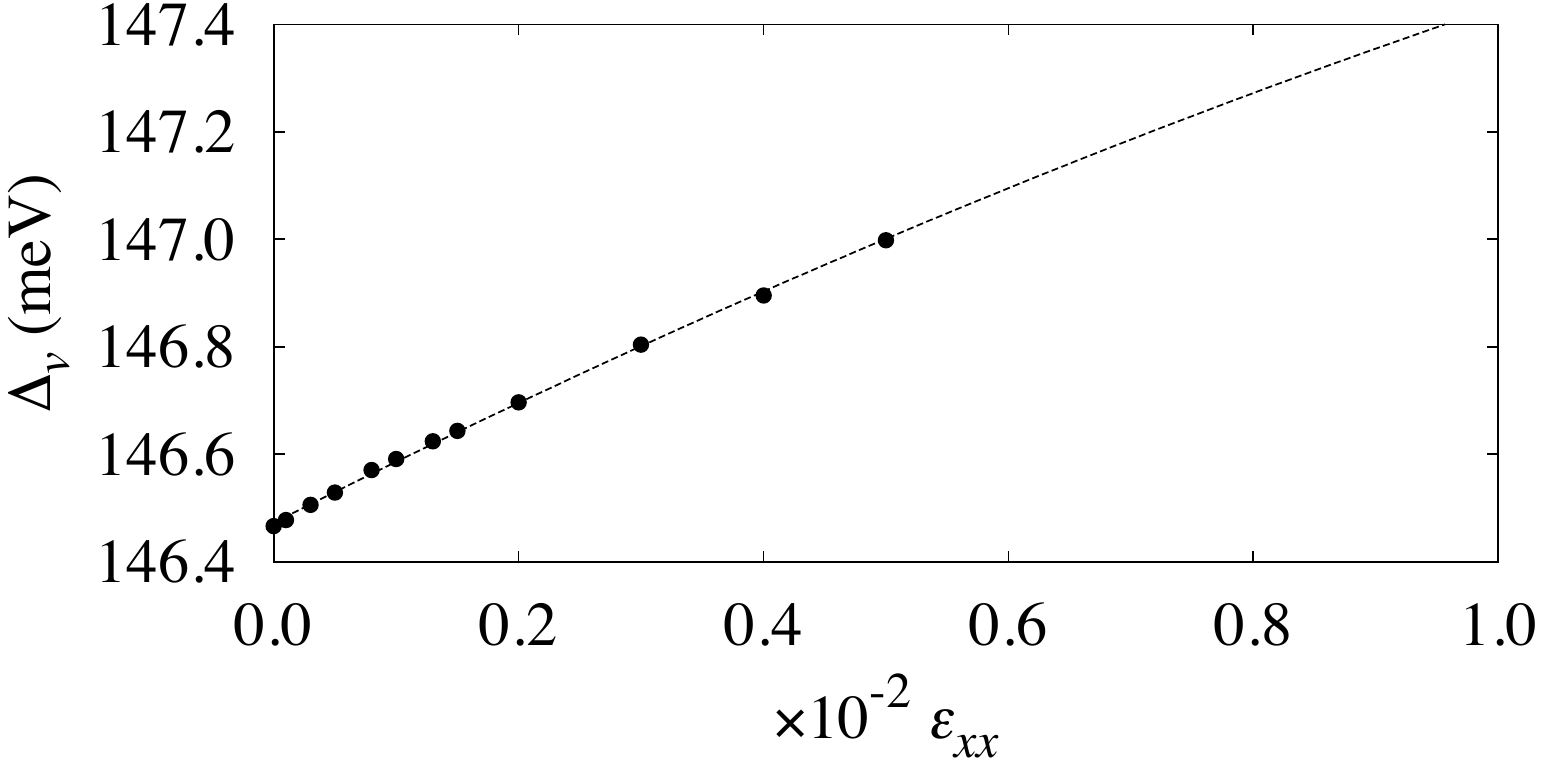}}
	
	\subfloat[]{\label{exx_ck}\includegraphics[width=75mm]{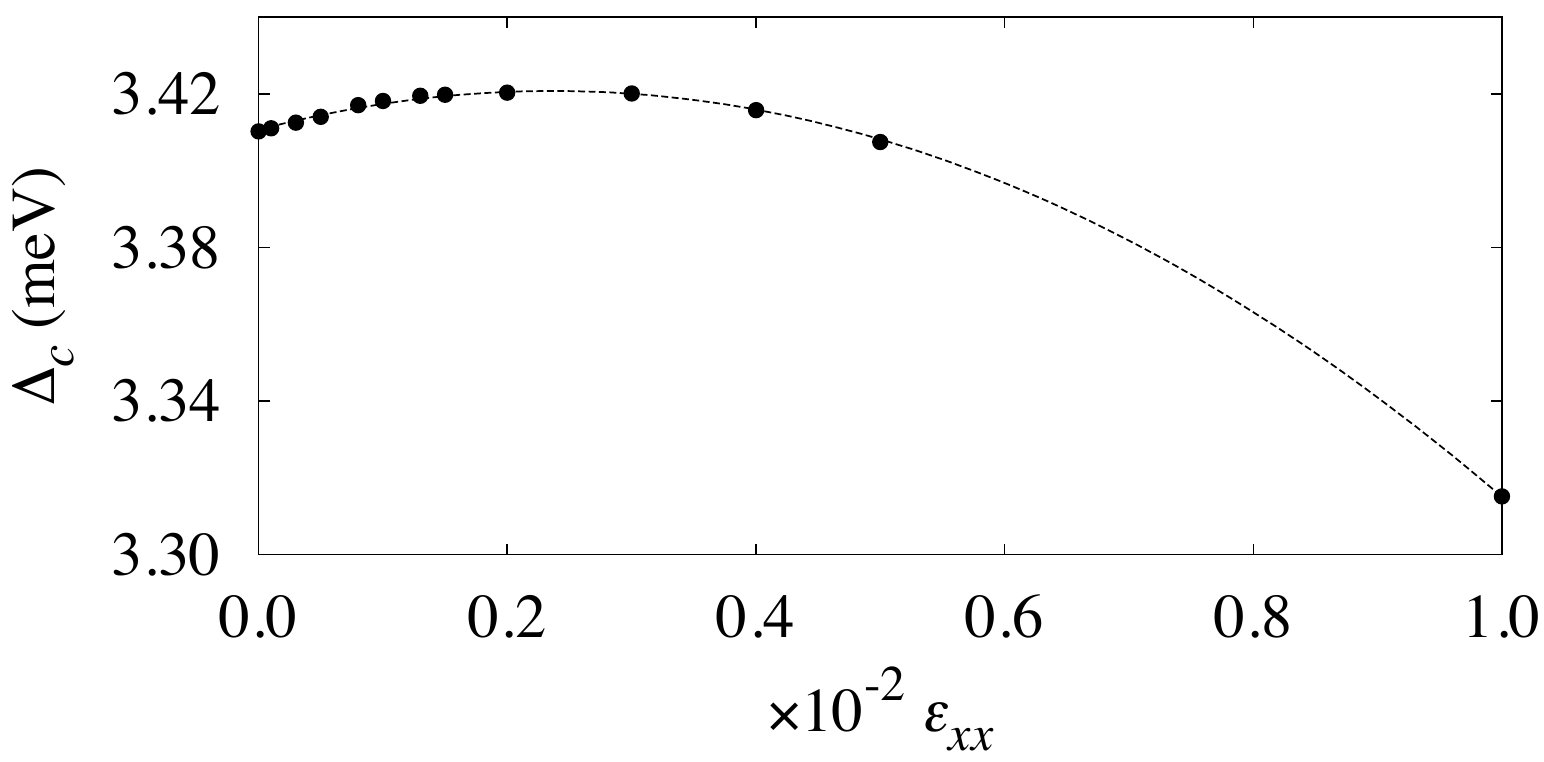}}
	
	\subfloat[]{\label{exx_gap}\includegraphics[width=75mm]{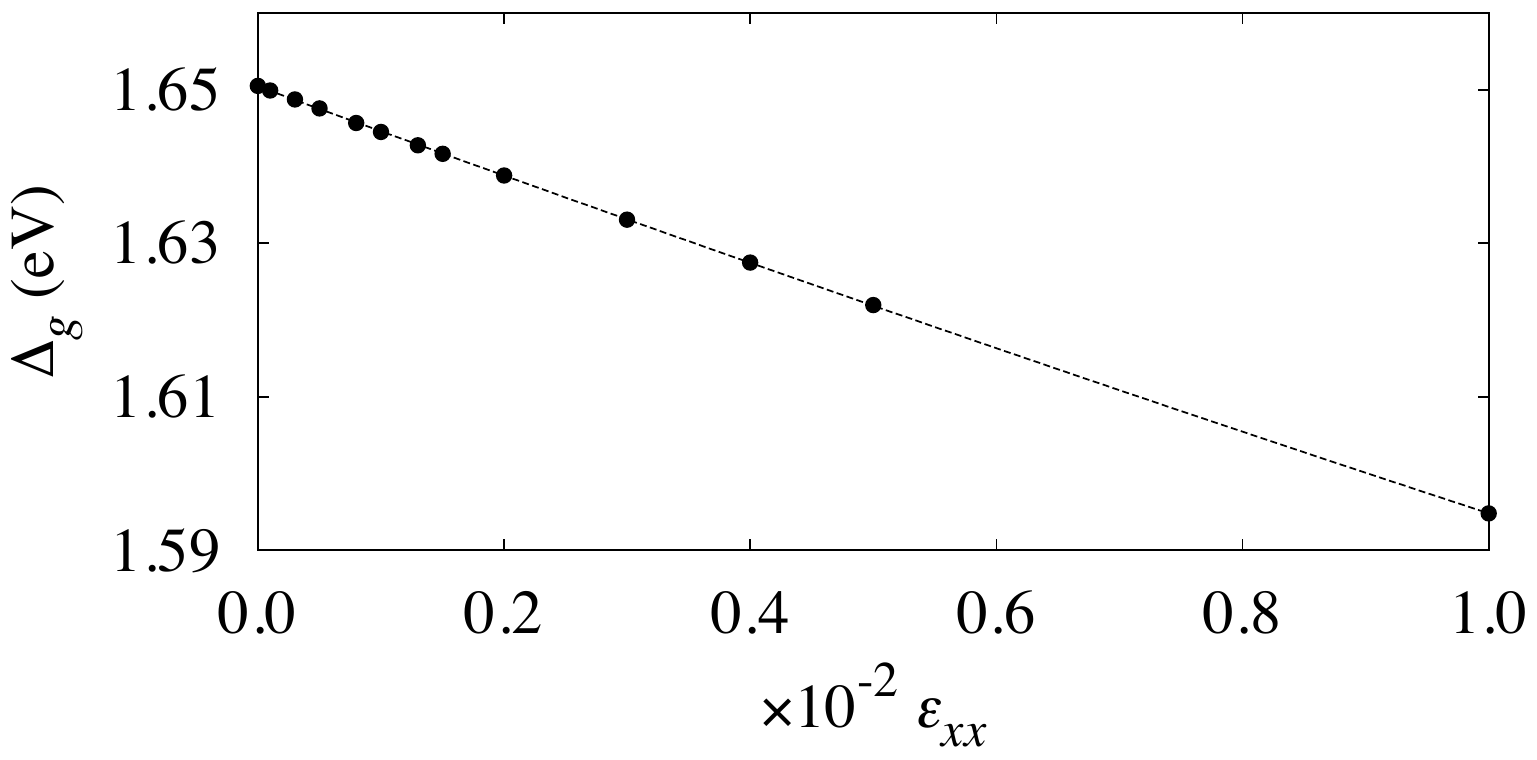}}
	\caption{(a) valence band splitting, (b) conduction band splitting, and (c) band gap of monolayer MoS$_2$ as a function of tensile strain ($\epsilon_{xx}$)}
	\label{exx}
\end{figure}
\begin{figure}[h]
	\subfloat[]{\label{exy_vk}\includegraphics[width=75mm]{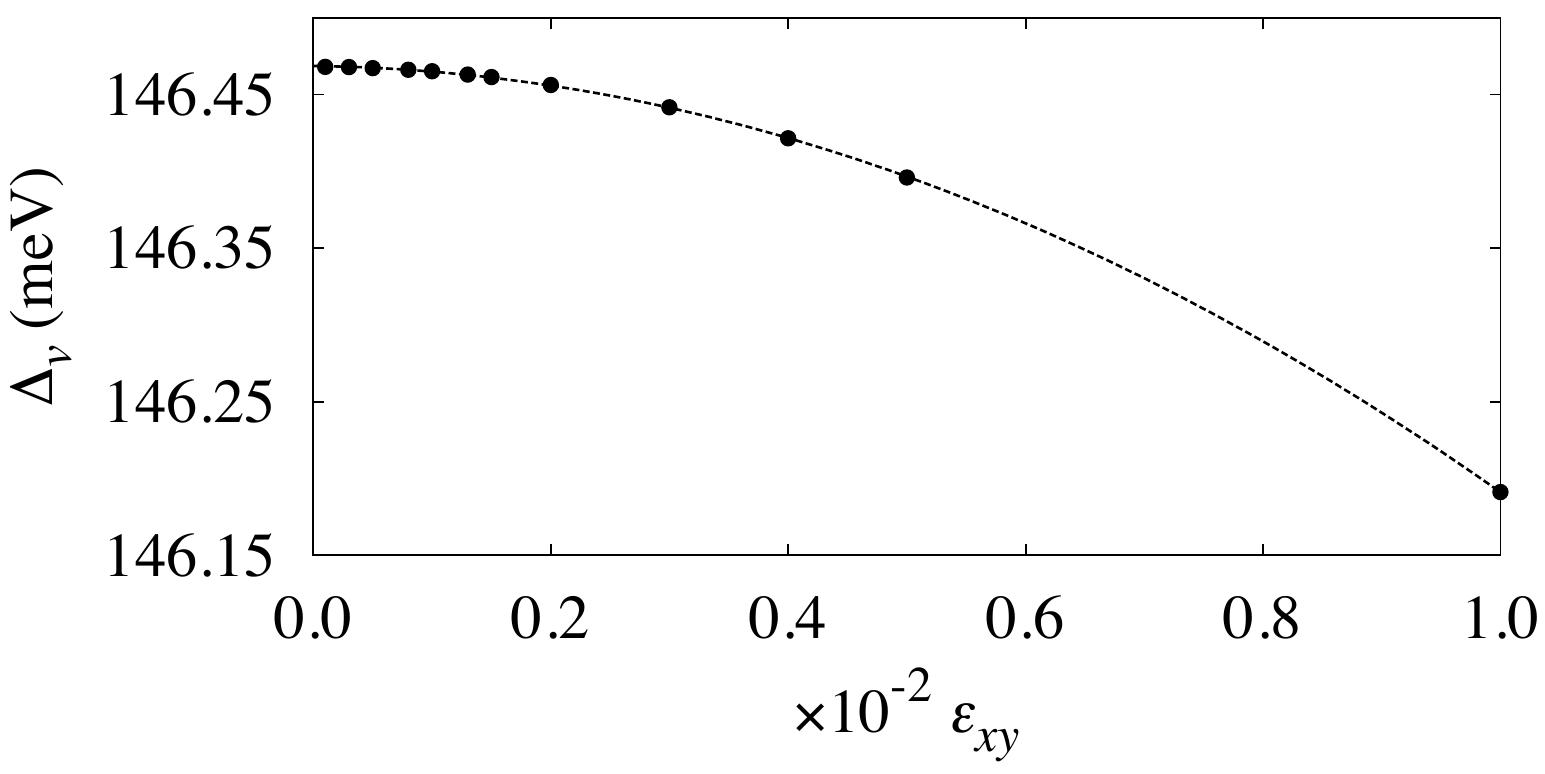}}
	
	\subfloat[]{\label{exy_ck}\includegraphics[width=75mm]{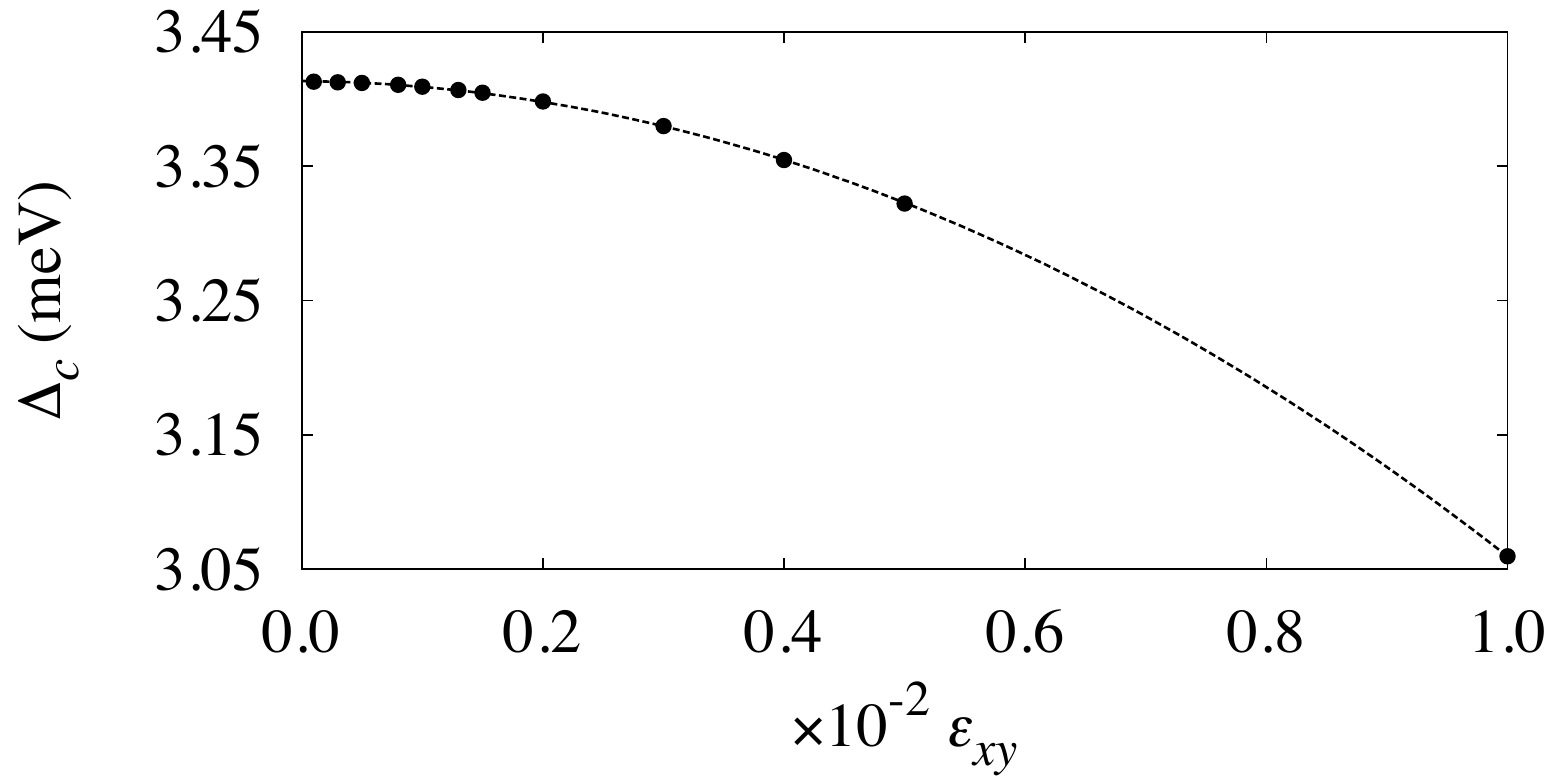}}
	
	\subfloat[]{\label{exy_gap}\includegraphics[width=75mm]{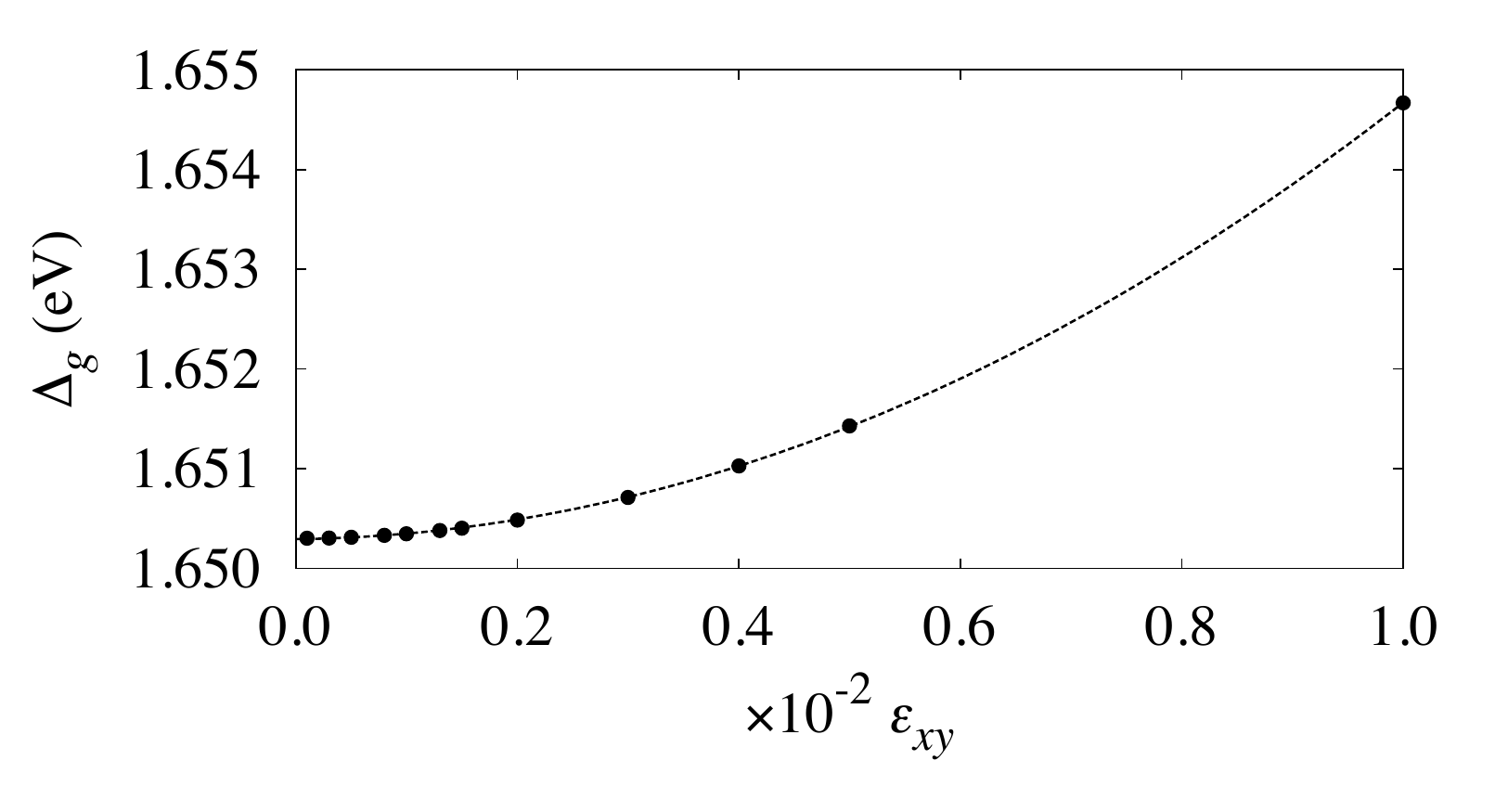}}
	\caption{(a) valence band splitting, (b) conduction band splitting, and (c) band gap of monolayer MoS$_2$ as a function of in-plane strain ($\epsilon_{xy}$)}
	\label{exy}
\end{figure}

Next, we consider the in-plane strains. As predicted in Eq.~(\ref{inplane}) for $\epsilon_{xy}$ with vanishing $b_1$, both VBM and CBM splittings are quadratic in $\epsilon_{xy}$. This behavior is shown in Figs.~\ref{exy_vk} and \ref{exy_ck}. By fitting Eq.~(\ref{inplane}) to the calculated data, $b_2$ parameters are found to be $-$2.67 eV and $-$3.45 eV for VBM and CBM, respectively. In case of tensile strain ($\epsilon_{xx}$), the splitting for both VBM and CBM show the combination of linear and quadratic in $\epsilon_{xx}$ characters. However, the positive linear coefficient ($b_1$) of VBM, which is 115 meV, is more than ten times larger than the one  for the CBM,  which is 8.85 meV. In contrast, their negative quadratic coefficients ($b_2$), $-$1.92~eV for the VBM and $-$1.81~eV for the CBM, are of the same order of magnitude. Therefore, the fit of the VBM splitting in Fig. \ref{exx_vk} is dominated by the linear term while the fit of CBM splitting in Fig. \ref{exx_ck} is quickly dominated by the quadratic term when $\epsilon_{xx}$ is larger than 0.001. Furthermore, the CBM splitting goes through a maximum as function of in-plane tensile strain.
All of the parameters mentioned above are tabulated in Table \ref{mos2_table}.

We also consider the gaps as function of strain in Figs. \ref{exz_gap}, \ref{exx_gap}, and \ref{exy_gap}. Band gaps as function of strain are fitted with a quadratic function
\begin{equation}
\Delta_g(\epsilon)=\Delta_g^0+c_1\epsilon+c_2\epsilon^2,
\end{equation}
and the resulting parameters are summarized in Table \ref{mos2_gap_table}. We can see that the tensile in-plane strain $\epsilon_{xx}$ has a much stronger effect on the band gap
than the shear strains. The linear decrease of the gap of about 59 meV/\% strain is close to the calculated value reported by Conley \textit{et al}.\cite{Conley}
The quadratic terms however are not negligible for the shear strains and  seem to be of similar magnitude for each strain although it is negative
for out of plane strain and positive for in-plane strain.

  \begin{table} 
  \caption{Parameters for monolayer MoS$_2$}
  \label{mos2_table}
 \begin{ruledtabular}
 \begin{tabular}{l c r r}
 Strain & Parameter & VBM & CBM \\
  \hline
 \multirow{2}{*} {$\epsilon_{xz}$,\,$\epsilon_{yz}$} & $a_1$ & 573 meV & 774 meV  \\
   		      & $b_2$     & $-$2.42 eV & 263 meV \vspace{1mm} \\
 \multirow{2}{*}{$\epsilon_{xx}$,\,$\epsilon_{yy}$} & $b_1$     & 115 meV & 8.58  meV \\
  		      & $b_2$     &  $-$1.92 eV & $-$1.81 eV \vspace{1mm} \\
$\epsilon_{xy}$ & $b_2$     & $-$2.67 eV & $-$3.45 eV \vspace{1mm}  \\	
 		      & $\Delta_0$ & 146 meV & 3.36 meV \\	
 \end{tabular}
 \end{ruledtabular}
 \end{table}

  \begin{table} 
  \caption{Parameters for the change in band gap of monolayer MoS$_2$}
  \label{mos2_gap_table}
 \begin{ruledtabular}
 \begin{tabular}{l c r }
 Strain & $c_1$ & $c_2$\\
  \hline
 $\epsilon_{xz}$ & $-175$ meV  & $-34.9$ eV  \\
 $\epsilon_{xx}$ & $-5.87$ eV    & 30.5 eV \\
 $\epsilon_{xy}$ & 13.5 meV   & 42.4 eV \\	
 \end{tabular}
 \end{ruledtabular}
 \end{table}

\subsection{Graphene}
\begin{figure}[h]
	\subfloat[]{\label{gra_exx}\includegraphics[width=75mm]{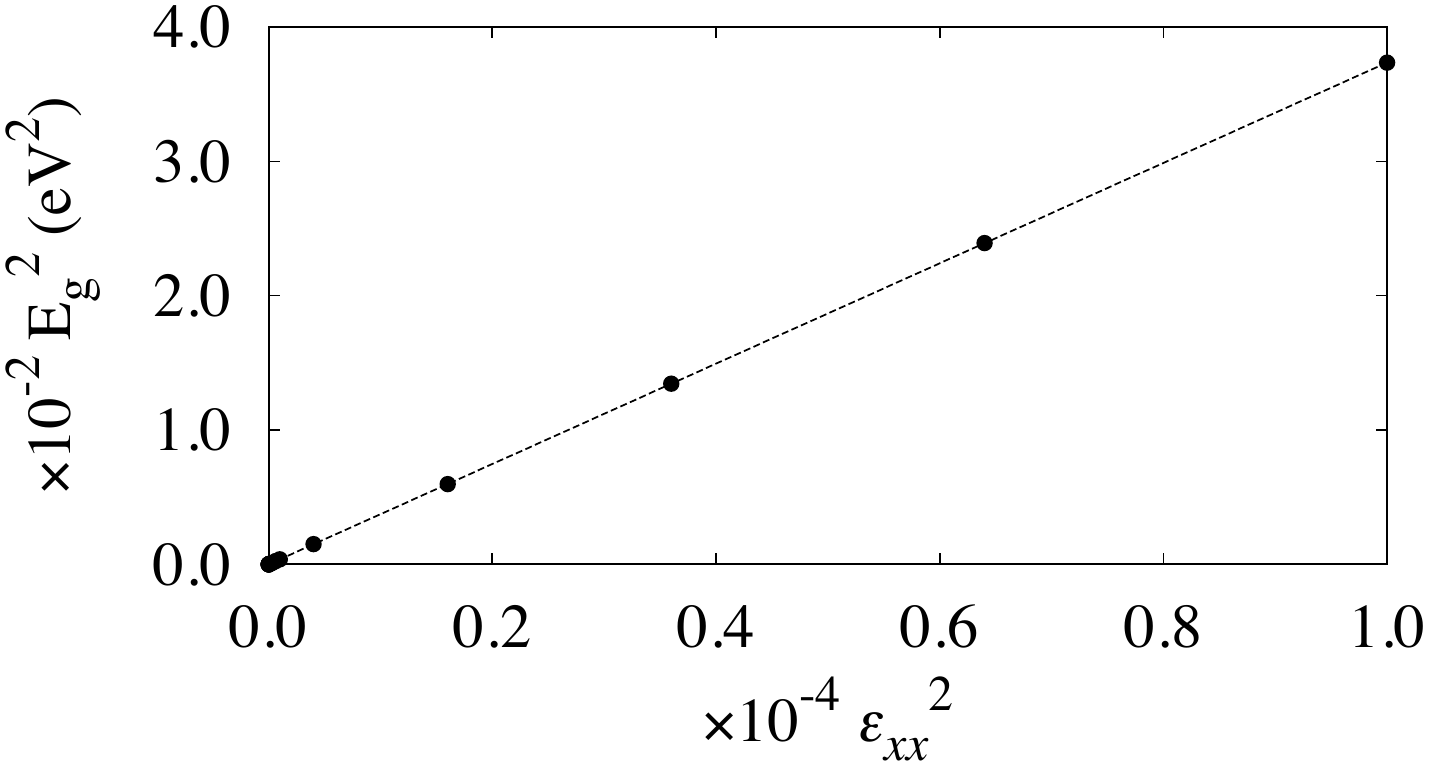}}
	
	
	\subfloat[]{\label{gra_eyz}\includegraphics[width=75mm]{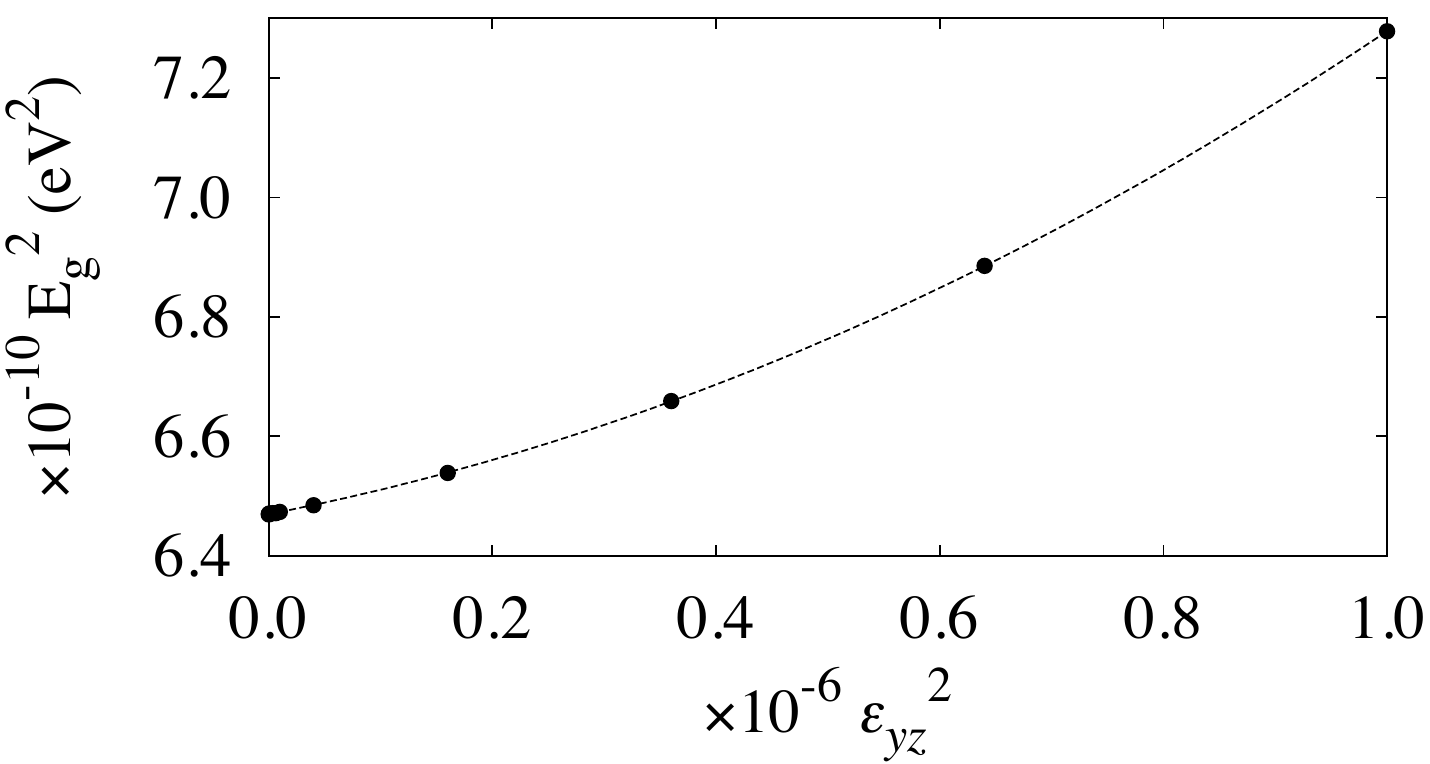}}
	\caption{Band gap squared with spin-orbit-coupling effect of graphene at $K$-point as function of (a) $\epsilon_{xx}$, and (b) $\epsilon_{yz}$ strain squared}
	\label{gra_fig}
\end{figure}

\begin{figure}[h]
	\subfloat[]{\label{exx_small}\includegraphics[width=75mm]{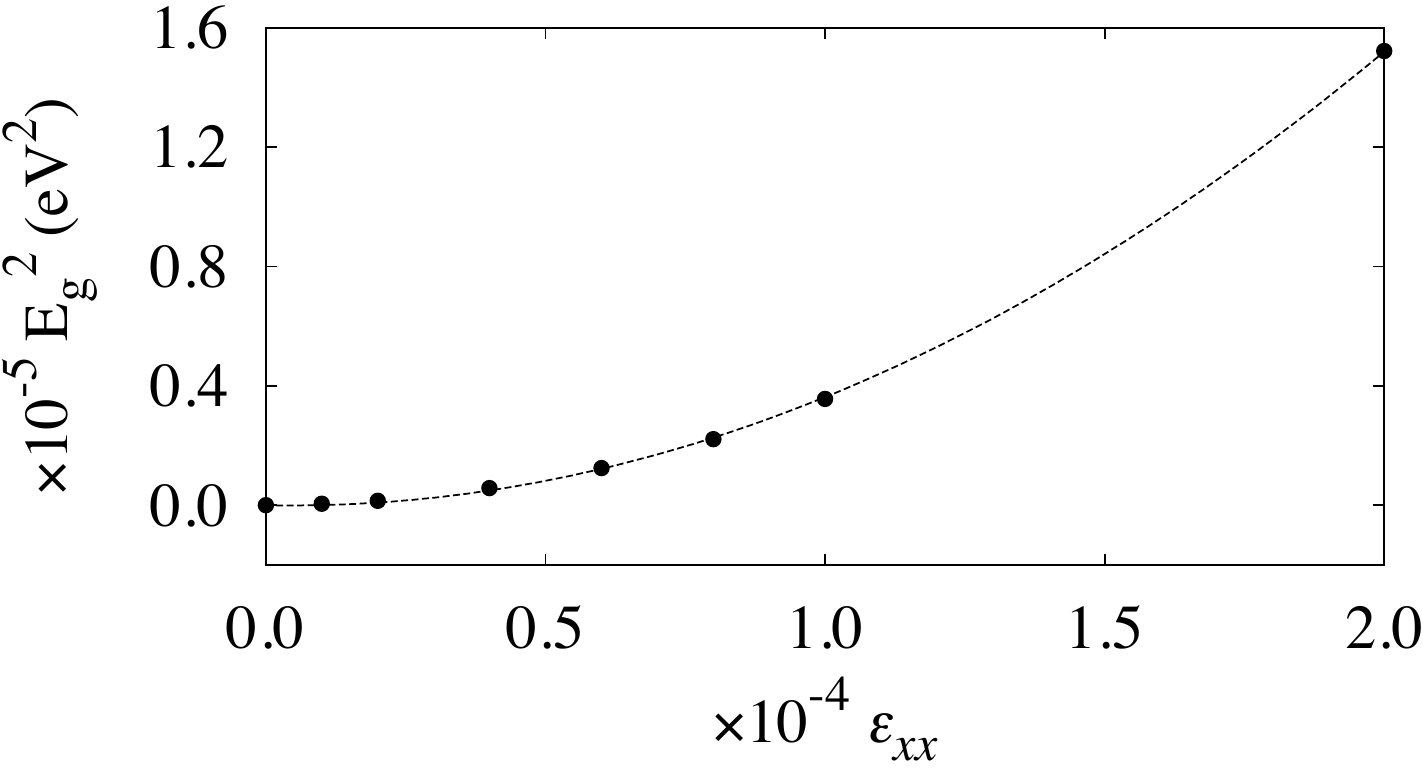}}

	\subfloat[]{\label{eyz_small}\includegraphics[width=75mm]{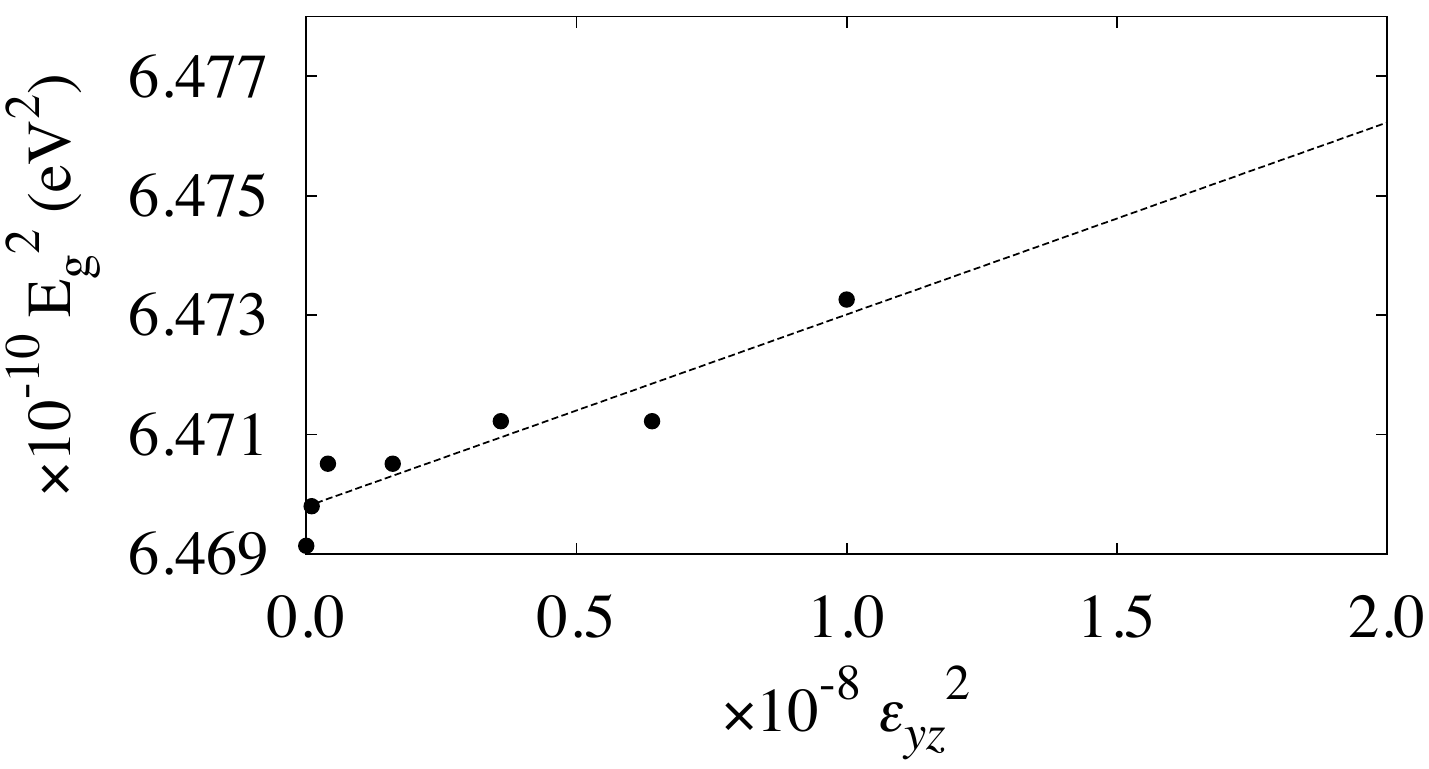}}
	\caption{Band gap squared of graphene at $K$-point as function of (a) $\epsilon_{xx}$, and (b) $\epsilon_{yz}^2$ for small strains.}
	\label{figsmallstrain}
\end{figure}

First, we note that we obtain a spin-orbit induced gap at the Dirac point
of about 26  $\mu$eV in the absence of strain. This result is in good agreement with
the linearized augmented plane wave (LAPW) calculations by Gmitra \textit{et al}.\cite{Gmitra09}
In Fig. \ref{gra_fig} we show the gap squared as function of strain squared for
in-plane and out-of-plane strain. Different in-plane strains,
$\epsilon_{xx}$, $\epsilon_{xx}-\epsilon_{yy}$, $\epsilon_{xy}$ gave somewhat different
results for the extracted fitting parameters because of numerical effects. We use these
to estimate the uncertainty on the extracted parameters, given in Table \ref{tabgraph}.
In a separate Figure \ref{figsmallstrain}
we show the small strain behavior as function
of $\epsilon_{xx}$ and as function of $\epsilon_{yz}^2$, which clearly shows that there is
no linear spin-orbit term for the in-plane strain case but only for the
out-of-plane strain case.
We note that for strains of the order of 0.001, the $c_{so}$ term is
comparable in magnitude with the $\Xi_o$ term. The linear fit in the small strain
region gives an uncertainty of about 8\% on the slope parameter or on $c_{so}$.

\begin{table}[h]
\caption{Parameters obtained from fittings for graphene.}
\begin{ruledtabular}
\begin{tabular}{c c c }
 $\Delta_D$ ($\mu$eV) & $\Xi_{o}$ (eV) & $c_{so}$ (meV)\\
\hline
 26$\pm$1 & 9.5$\pm0.5$ & 2.9$\pm0.3$   \\
\end{tabular}
\end{ruledtabular} \label{tabgraph}
\end{table}

\section{Conclusions}

The strain-dependent spin splittings of MoS$_2$ and graphene are systematically studied.
The spin-orbit induced splittings of the valence-band maximum, conduction-band minimum and the gap in MoS$_2$ as well as the spin-orbit induced gap
in graphene were studied as function of strain using first-principles calculations.
Various spin-dependent and spin-independent deformation potentials were extracted, in comparison with strain Hamiltonians governed by the method of invariants. The importance of these deformation potentials is that they directly dictate the strength of intrinsic spin-flip and momentum scattering due to electron-phonon interaction. The obtained spin-dependent deformation potential for graphene $c_{so}\approx 3$ meV, rather than the much smaller Dirac gap ($\sim$26~$\mu$eV), is the  important  parameter in studying the intrinsic spin relaxation in graphene. Similarly, sizable spin-dependent deformation potentials on the order of atomic spin-orbit coupling ($\gtrsim$$\,$0.1~eV) render strong electron spin relaxation in MoS$_2$, despite the tiny spin splitting in its conduction band minimum ($\sim$3.4~meV).


\acknowledgments{The calculations made use of the
High Performance Computing Resource in the Core Facility
for Advanced Research Computing at Case Western Reserve University.
The work at CWRU was supported by NSF under grant DMR-1104595. The work at the University of Rochester was supported by NRI-NSF,
NSF, and DTRA Contracts No. DMR-1124601, No. ECCS-1231570, and No. HDTRA1-13-1-0013, respectively.}

\appendix

\section{Detailed derivation of in plane spin-dependent deformations  of M\lowercase{o}S$_2$}\label{app:b1}

The expressions of  $b_1$ in Eq.~(\ref{inplane}) are obtained through perturbation theory and method of invariants. The general perturbation brought by in-plane strain is
\begin{eqnarray}
H^{in}_{strain} &=& \sum_{i,j=\{x,y\}}  r_i \frac{\partial V}{\partial r_j} \epsilon_{ij}\nonumber\\
 &=& \frac{1}{2} \left(x\frac{\partial V}{\partial x} + y\frac{\partial V}{\partial y}\right)(\epsilon_{xx}+\epsilon_{yy})\nonumber\\
  &&+ \frac{1}{2} \left(x\frac{\partial V}{\partial y} - y\frac{\partial V}{\partial x}\right)(\epsilon_{xy}-\epsilon_{yx})\nonumber\\
  &&+ \frac{1}{4} \left[\left(x\frac{\partial V}{\partial x} - y\frac{\partial V}{\partial y}\right) + i \left(x\frac{\partial V}{\partial y} + y\frac{\partial V}{\partial x}\right) \right]\nonumber\\
  &&\times[(\epsilon_{xx}-\epsilon_{yy})-i(\epsilon_{xy}+\epsilon_{yx})]\nonumber\\
  &&+ \frac{1}{4} \left[\left(x\frac{\partial V}{\partial x} - y\frac{\partial V}{\partial y}\right) - i \left(x\frac{\partial V}{\partial y} + y\frac{\partial V}{\partial x}\right) \right]\nonumber\\
  &&\times[(\epsilon_{xx}-\epsilon_{yy})+i(\epsilon_{xy}+\epsilon_{yx})],\label{eq:inplane_strain_Hamilt}
\end{eqnarray}
where the first two terms both transform as $K_1$ which is the identity irreducible representation (IR) of the $K$-point $C_{3h}$ group.\cite{Song_PRL13} The last two terms transform as $K_{2,3}$ IRs. For simplicity, we have omitted the $p_i p_j \epsilon_{ij}/m$ terms in $H^{in}_{strain}$ (they transform the same as $r_i \partial V/\partial r_j$). Using the one-dimension nature of all the IRs in $C_{3h}$ ($K_i \times K_i^{\ast} = K_1$), intra-band coupling via the invariants $(\epsilon_{xx}+\epsilon_{yy})\sigma_z$ or $(\epsilon_{xy}-\epsilon_{yx})\sigma_z$ is symmetry-allowed in all of the bands. The corresponding integral constant of this intra-band coupling comes from second-order perturbation (one due to strain and another due to spin-orbit) as well as from first-order perturbation (strain-modified spin-orbit interaction). Below, we show that the two parts cancel-out partially for a general integral constant associated with $\epsilon_{ij}\sigma_z$ of spin-independent band $K_n$. The first-order perturbation part due to strain-modified spin-orbit interaction reads
\begin{eqnarray}
&&\lambda\langle K_{n} |[\bm\nabla (r_i\frac{\partial V}{\partial r_j})\times \mathbf{p}]_z | K_{n}\rangle \nonumber\\
&&+\lambda\langle K_{n} |- \delta_{xi}\frac{\partial V}{\partial r_j}p_y + \delta_{yi}\frac{\partial V}{\partial r_j}p_x| K_{n}\rangle\nonumber\\
&&+\lambda\langle K_{n} |- \delta_{yi}\frac{\partial V}{\partial x}p_j + \delta_{xi}\frac{\partial V}{\partial y}p_j| K_{n}\rangle\nonumber\\
&=&\lambda\langle K_{n} |r_i\frac{\partial(\bm\nabla  V\times \mathbf{p})_z }{\partial r_j}| K_{n}\rangle + \nonumber\\
&&\lambda\langle K_{n} |(- \delta_{yi}\frac{\partial V}{\partial x} + \delta_{xi}\frac{\partial V}{\partial y})p_j| K_{n}\rangle,\label{eq:first-order-perturb}
\end{eqnarray}
where $\lambda=\hbar/4m_0^2 c^2$. Eq.~(\ref{eq:first-order-perturb}) assumes rigid-ion approximation, but it retains the correct symmetry and does not affect the result qualitatively. The second-order perturbation part reads,
\begin{eqnarray}
 \frac{\!\!2{\rm Re}\!\left[\langle K_{n} |\!\left(\!r_i\frac{\partial V}{\partial r_j}\!+\!\frac{p_i p_j}{m}\!\right)\!\! {\displaystyle \sum_{n'}}| K_{n'}\rangle \langle K_{n'} | \lambda(\bm\nabla V\!\times \!\mathbf{p})_z | K_{n}\rangle \right]} {E_{K_{n}}-E_{K_{n'}}},\!\!\!\!\!\!\!\!\!\!\nonumber\\  \!\!\!\!\!\!\!\!\!\! \label{eq:second-order-perturb}
\end{eqnarray}
where $K_{n'}$ is other bands at $K$ point excluding the $K_{n}$ band. Using the fact that $\frac{\partial V}{\partial r_j} =[p_j,\!H]$, we get that
\begin{eqnarray}
&&\!\!\frac{2{\rm Re}\left[\langle K_{n} |r_i\frac{\partial V}{\partial r_j} \sum_{n'}| K_{n'}\rangle \langle K_{n'} | \lambda(\bm\nabla V\times \mathbf{p})_z | K_{n}\rangle \right]} {E_{K_{n}}-E_{K_{n'}}}
\nonumber\\
&=&\frac{2}{\hbar}{\rm Im}\left[\langle K_{n} |r_i p_j \sum_{n'}| K_{n'}\rangle \langle K_{n'} | \lambda(\bm\nabla V\times \mathbf{p})_z | K_{n}\rangle \right]. \nonumber\\
\end{eqnarray}
Dispensing with the identity operator altogether, we get
\begin{eqnarray}
&&-\frac{i}{\hbar}\langle K_{n} |r_i p_j  \lambda(\bm\nabla V\times \mathbf{p})_z - \lambda(\bm\nabla V\times \mathbf{p})_z r_i p_j| K_{n}\rangle
\nonumber\\
&=&-\lambda\langle K_{n} |r_i \frac{ \partial(\bm\nabla V\times \mathbf{p})_z}{\partial r_j}| K_{n}\rangle \nonumber\\
&&+ \lambda\langle K_{n} |( \delta_{yi}\frac{\partial V}{\partial x}- \delta_{xi}\frac{\partial V}{\partial y}) p_j| K_{n}\rangle.
\end{eqnarray}
This term cancels out the first-order perturbation part [Eq.~(\ref{eq:first-order-perturb})], and the net result from adding Eqs.~(\ref{eq:first-order-perturb}) and (\ref{eq:second-order-perturb}) is
\begin{eqnarray}
\frac{2{\rm Re}\left[\langle K_{n} |\frac{p_i p_j}{m} \sum_{n'}| K_{n'}\rangle \langle K_{n'} | \lambda(\bm\nabla V\times \mathbf{p})_z | K_{n}\rangle \right]} {E_{K_{n}}-E_{K_{n'}}}.\label{eq:net-result}
\end{eqnarray}
Therefore, by Eq.~(\ref{eq:inplane_strain_Hamilt}) and the argument that followed it, the integral constant $b_{1,z}$ associated with shear strain drops ($p_yp_x-p_xp_y=0$), while the integral constant associated with dilation strain remains ($p_xp_x+p_yp_y\neq 0$). In practice, $\epsilon_{xy}-\epsilon_{yx}$ is treated as 0 since rotation does not induce energy perturbation, and the accompanied deformation potentials are not evaluated. The above derivation justifies this practice from the detailed interaction Hamiltonian, while Eq.~(\ref{eq:net-result}) times two gives the detailed expression for $b_1$ with $i=j$.

Finally, when $|K_{n}\rangle$ is dominated by orbitals with zero angular momentum (e.g., lowest conduction band of MoS$_2$), the majority of $b_1$ vanishes, since $(\bm\nabla V\times \mathbf{p})_z| d_{z^2}\rangle=0$. So $b_{1}$ is small in this case and is proportional to  the small components from S $p_x,p_y$ orbitals,\cite{Zhu_PRB11} and the small deviation of the Mo atomic potential from spherical symmetry due to the crystal structure.\cite{Song_PRL13}

\begin{widetext}

\section{Character tables and invariants}\label{app:b2}

\begin{table}[!htbp]
{ 
\caption{Character table of the $D_{3h}$ point double group.\cite{Bradley_Cracknell72} It is used to construct the strain-dependent Hamiltonian in the $K$~point of graphene. $\{x,y,z\}$ and $\{R_x,R_y,R_z\}$ represent components of a polar and axial vectors, respectively. The  $x$-axis is defined along the zigzag edge direction and the $y$-axis is along the armchair direction. A strain tensor component $\epsilon_{ij}$ transforms as the product of the $i$-\textit{th} and $j$-\textit{th} components of a polar-vector.} \label{tab:CharactorTable_D_3h} \renewcommand{\arraystretch}{1.5} \tabcolsep=0.03cm
\begin{tabular}{cc|ccc ccc ccc cc}

\hline \hline
\multicolumn{2}{c|}{$D_{3h}$} &$E$& $\bar{E}$& $C^+_3\,C^-_3$& $\bar{C}^+_3\,\bar{C}^-_3$  &$\sigma_h\,\bar{\sigma}_h$ &$S^+_3\,S^-_3$ & $\bar{S}^+_3\,\bar{S}^-_3$ & $C'_{2i} \bar{C}'_{2i}$ & $\sigma_{vi} \bar{\sigma}_{vi}$ & invariants  \\ \hline
$A'_1$ &$\Gamma_1$            & 1 & 1        &1              &1
                & 1                          & 1            &1
              &1                        &  1 & $z^2, x^2+y^2$, $x(3y^2-x^2)$
  \\
$A'_2$ &$\Gamma_2$            & 1 & 1        &1              &1
                & 1                          & 1            &1
              &$-1$    &$-1$ & $R_z,\,y(3x^2-y^2)$
  \\
$A''_1$ &$\Gamma_3$           & 1 & 1        &1              &1
                &$-1$                        &$-1$          &$-1$
               &1                        &$-1$     \\
$A''_2$ &$\Gamma_4$           & 1 & 1        &1              &1
                &$-1$                        &$-1$          &$-1$
               &$-1$                        &$1$ & $z$
  \\
$E''$ &$\Gamma_5$             & 2 & 2        &$-1$           &$-1$
                &$-2$                        &1             &1
               &0                        &0  & $\{R_x,R_y\},\,\{yz,-xz\}$
\\
$E'$ &$\Gamma_6$              & 2 & 2        &$-1$           &$-1$
                &$2$                        &$-1$            &$-1$
               &0                        &0 & $\{x,y\},\,\{ x^2-y^2,
-2xy\}$
\\
$\bar{E}_1$ &$\Gamma_7$       & 2 &$-2$        &$1$         &$-1$
               &0                        &$\sqrt{3}$         &$-\sqrt{3}$
               &0                        &0
&$\{\uparrow_z,\downarrow_z\},\{iR_x\!-\!R_y\!\downarrow_z,iR_x\!+\!R_y\!\uparrow_z\!\}$
 \\
$\bar{E}_2$ &$\Gamma_8$       & 2 &$-2$        &$1$         &$-1$
               &0                        &$-\sqrt{3}$         &$\sqrt{3}$
               &0                        &0
    \\
$\bar{E}_3$ &$\Gamma_9$       & 2 &$-2$        &$-2$         &$2$
               &0                        &0                  &$0$
               &0                        &0     &
$\{iR_x\!+\!R_y\!\downarrow_z,iR_x\!-\!R_y\!\uparrow_z\!\}$\\
\hline\hline
\end{tabular} }
\end{table}

\begin{table}[!htbp]
{
\caption{Character table of the $C_{3h}$ point double group.\cite{Bradley_Cracknell72} It is used to construct the strain-dependent Hamiltonian in the $K$~point of MX$_2$. The function notations are defined in the same way as in Table~\ref{tab:CharactorTable_D_3h}.} \label{tab:CharactorTable_K}
\renewcommand{\arraystretch}{1.5} \tabcolsep=0.03cm \begin{tabular}{cc|ccc ccc ccc ccc c} \hline \hline \multicolumn{2}{c|}{$C_{3h}$} &$E$& $C^+_3$& $C^-_3$ &$\sigma_h$ &$S^+_3$ & $S^-_3$ & $\bar{E}$& $\bar{C}^+_3$& $\bar{C}^-_3$ &$\bar{\sigma}_h$ & $\bar{S}^+_3$ & $\bar{S}^-_3$  & invariants \\ \hline
$A'$ &$K_1$                   & 1 & 1      &1        &1          & 1
& 1       &1         &1             &  1            &  1              &  1
           &  1  &  $R_z,x^2+y^2,z^2$ \\
$^2 E'$ &$K_2$                & 1 &$\omega$&$\omega^*$ &1
&$\omega$&$\omega^*$ & 1      &$\omega$      &$\omega^*$     &1
    &$\omega$       &$\omega^*$  &$x - iy,\,2xy-i(x^2-y^2)$ \\
$^1 E'$ &$K_3$                & 1 &$\omega^*$&$\omega$ &1
&$\omega^*$&$\omega$ & 1      &$\omega^*$    &$\omega$       &1
    &$\omega^*$     &$\omega$ &$x + iy,\,2xy+i(x^2-y^2)$ \\
$A''$ &$K_4$                  & 1 & 1      &1        &$-1$       &$-1$
& $-1$      & 1      & 1            &1              &$-1$
&$-1$           & $-1$   & $z$  \\
$^2 E''$ &$K_5$               & 1 &$\omega$&$\omega^*$ &$-1$
&$-\omega$&$-\omega^*$ & 1    &$\omega$      &$\omega^*$     &$-1$
      &$-\omega$      &$-\omega^*$ &$R_x - iR_y,\,yz+ixz$
\\
$^1 E''$ &$K_6$               & 1 &$\omega^*$&$\omega$ &$-1$
&$-\omega^*$&$-\omega$ & 1    &$\omega^*$    &$\omega$       &$-1$
      &$-\omega^*$    &$-\omega$  &$R_x + iR_y,\,yz-ixz$ \\
$^1\bar{E}_3$ &$K_7$          & 1 &$-\omega$&$-\omega^*$&$i$
&$-i\omega$&$i\omega^*$ & $-1$   &$\omega$      &$\omega^*$     &$-i$
       &$i\omega$      &$-i\omega^*$ & $\uparrow_z$\\
$^2\bar{E}_3$ &$K_8$          & 1 &$-\omega^*$&$-\omega$&$-i$
&$i\omega^*$&$-i\omega$ & $-1$   &$\omega^*$   &$\omega$       &$i$
      &$-i\omega^*$   &$i\omega$ & $\downarrow_z$\\
$^2\bar{E}_2$ &$K_9$          & 1 &$-\omega$&$-\omega^*$&$-i$
&$i\omega$  &$-i\omega^*$ & $-1$ &$\omega$     &$\omega^*$     &$i$
      &$-i\omega$     &$i\omega^*$   \\
$^1\bar{E}_2$ &$K_{10}$       & 1 &$-\omega^*$&$-\omega$&$i$
&$-i\omega^*$ &$i\omega$  & $-1$ &$\omega^*$    &$\omega$       &$-i$
       &$i\omega^*$    &$-i\omega$\\
$^1\bar{E}_1$ &$K_{11}$       & 1 &$-1$    &$-1$     &$i$        &$-i$
&$i$       & $-1$      &$1$           &$1$            &$-i$
&$i$            &$-i$       \\
$^2\bar{E}_1$ &$K_{12}$       & 1 &$-1$    &$-1$     &$-i$       &$i$
&$-i$       & $-1$      &$1$           &$1$            &$i$
&$-i$            &$i$      \\
\hline\hline
\end{tabular} }
\footnotetext{$\qquad \qquad \qquad \qquad \,\,$$\omega=\exp(2\pi i/3)$. } \end{table}

\end{widetext}

\bibliography{mos2,lmto,gw}

\end{document}